\documentclass[useAMS,usenatbib,usegraphicx]{mn2e}

\usepackage{amssymb}
\usepackage{graphicx}
\usepackage{wasysym}
\usepackage{pifont}
\newcommand{\re}{R_{\textrm{\tiny{e}}}}
\newcommand{\rl}{R_{\textrm{\tiny{lens}}}}

\newcommand{\ml}{M_{\textrm{\tiny{lens}}}}
\newcommand{\ms}{M_{\textrm{\tiny{stel}}}}
\newcommand{\msun}{M_\odot}
\newcommand{\mv}{M_{\textrm{\tiny{vir}}}}
\newcommand{\mh}{M_{\textrm{\tiny{halo}}}}
\newcommand{\msph}{M_{\textrm{\tiny{sph}}}}
\newcommand{\rvam}{r_{\textrm{\tiny{vir}}}^{\textrm{\tiny{AM}}}}
\newcommand{\mvam}{M_{\textrm{\tiny{vir}}}^{\textrm{\tiny{AM}}}}
\newcommand{\cinit}{c_{\textrm{\tiny{init}}}}
\newcommand{\cfinal}{c_{\textrm{\tiny{final}}}}
\newcommand{\rv}{r_{\textrm{\tiny{vir}}}}
\newcommand{\rs}{r_{\textrm{\tiny{S}}}}
\newcommand{\lfsf}{LFSF}

\title{Diagnostics of Baryonic Cooling in Lensing Galaxies}

\author[D. Leier, I. Ferreras, P. Saha]{Dominik Leier$^{1}$\thanks{Email: \tt leier@ari.uni-heidelberg.de}, Ignacio Ferreras$^{2}$, Prasenjit Saha$^{3}$
 \\
$^{1}$Astronomisches Rechen-Institut, Zentrum f\"ur Astronomie der
 Universit\"at Heidelberg, M\"onchhofstrasse 12-14, 69120 Heidelberg,
 Germany \\
$^{2}$Mullard Space Science Laboratory, University College London,
 Holmbury St Mary, Dorking, Surrey RH5 6NT, UK\\
$^{3}$Institute for Theoretical Physics, University of Z\"urich,
 Winterthurerstrasse 190, CH-8057 Z\"urich, Switzerland}

\begin{document}

\date{Accepted 2012 April 24. Received 2012 February 27; in original form 2011 September 26.}

\maketitle

\begin{abstract}
  Theoretical studies of structure formation find an inverse
  proportionality between the concentration of dark matter haloes and
  virial mass. This trend has been recently confirmed for $\mv\gtrsim
  6\times 10^{12}\msun$ by the observation of the X-ray emission from
  the hot halo gas. We present an alternative approach to this
  problem, exploring the concentration of dark matter haloes over
  galaxy scales on a sample of 18 early-type systems. Our $c-\mv$
  relation is consistent with the X-ray analysis, extending towards
  lower virial masses, covering the range from $~4\times 10^{11}\msun$
  up to $5\times 10^{12}\msun$. A combination of the lensing analysis
  along with photometric data allows us to constrain the baryon
  fraction within a few effective radii, which is compared with
  prescriptions for adiabatic contraction (AC) of the dark matter
  haloes. We find that the standard methods for AC are strongly
  disfavored, requiring additional mechanisms -- such as mass loss
  during the contraction process -- to play a role during the
  phases following the collapse of the haloes.
\end{abstract}


\section{Introduction}

Dark matter haloes constitute the scaffolding on which the luminous
component of cosmic structure can be detected in the form of galaxies.
The connection between ordinary matter (i.e. ``baryons'') and the
dominant dark matter can only be found indirectly because of the
elusive nature of the latter. Theoretical studies give useful insight
under a number of assumptions about the properties of the dark matter
component. N-body simulations predict a universal density profile
\citep{nfw96} driven by two parameters, the mass of the halo --
usually defined out to a virial radius -- and the concentration, given
as the ratio between the scale length of the halo ($\rs$, where
$\Delta\log\rho/\Delta\log r=-2$) and the virial radius. A number of
N-body simulations \citep[see e.g.][]{BKS01,neto07,macc07} display a
significant trend between these two parameters, mainly driven by the
hierarchical buildup of structure. Massive haloes assemble at later
epochs -- when the background density is lower because of the
expansion of the Universe. Hence, we expect a trend whereby
concentration decreases with increasing galaxy mass. An observational
confirmation of this theoretical result by means of model-independent
mass reconstruction gives insights in the interplay between luminous
and dark matter.

Over cluster scales it is possible to explore the dark matter halo via
its effect on the gravitational potential. The X-ray bremsstrahlung
emission from the hot gas in the intracluster medium acts as a tracer
of the potential. If assumptions are made about the dynamical state of
the cluster, one can constrain the halo properties \citep[see
  e.g.][]{Sato00}. \citet{B07} found a significant correlation between
dark matter halo and mass, as expected from theoretical studies. Their
sample covers a range from massive early-type galaxies up to galaxy
clusters.  In this paper, we extend the observational effort towards
lower masses, constraining the haloes over galaxy scales by the use of
strong gravitational lensing. By targeting a sample of strong lenses
at moderate redshift (z$\sim$0.5) we probe the mass distribution out
to a few ($\sim$4) effective radii. Other approaches to probe
  dark matter haloes over galaxy scales involve the use of dynamical
  tracers such as the bulk of the stellar populations
  \citep{Gerhard01}. While Sauron data, as used in \cite{cap06} is
  restricted to 1--2 effective radii by the surface brightness
  detection limit of the observations and contains thus little
  information on the properties of the halos, studies based on more
  extended data are available \citep[see
    e.g.][]{Thomas09}. Additionally, planetary nebul\ae\ in the outer
  regions of galaxies \citep{Hui1995,rom03,Deason2011} or globular
  clusters \citep{Cote2001,Romanowsky2009,schub10} make it possible
  to probe the dark matter profile. Being evolved phases of the underlying stellar
populations, planetary nebul\ae\ can be considered unbiased tracers of
the gravitational potential \citep[see e.g.][]{coccato09}. 
Through their emission lines, it is possible to trace their kinematics
out to large distances, reaching out to $\sim 5\re$\citep{Douglas2002}.  
However, the interpretation of the results is difficult because of the
uncertainties regarding the parameterisation of the halo mass,
anisotropy, shape, or inclination \citep{delorenzi09}. Gravitational
lensing studies do not suffer from the inherent degeneracies of
methods regarding the modelling of the dynamical tracers,
although it is fair to say that
lensing studies have other modelling degeneracies, as we discuss in
the following Section. Ultimately, a comparison between all these
methods is key to a robust assessment of the $c-\mv$ relation.

Our recent study of stellar and total mass in lensing galaxies
\citep[][hereafter \lfsf]{Leier2011} indicated an inverse trend of
concentration with mass.  Those results, however, applied to radii
much smaller than the virial radius $\rv$. In this work, we extend
this analysis by extrapolating the inferred dark-matter profiles out
to $\rv$, to determine whether the concentration/mass trend persists to lower
masses, i.e., galaxy haloes.  We then try to reconstruct the possible
concentrations of the dark-matter haloes before adiabatic contraction
due to the baryons.

We consider a sample of 18 early-type lensing galaxies.  In
\lfsf\ these galaxies (plus two disk galaxies and one ongoing
major merger) were decomposed into stellar and dark-matter profiles.
The stellar-mass or $\ms$ profiles were obtained by fitting stellar
population synthesis models to the star light. The lensing-mass or
$\ml$ profiles were obtained from lens models.  The difference between
these is assumed to be the dark-matter profile.

\begin{figure*}
\begin{center}
\includegraphics[scale=0.7]{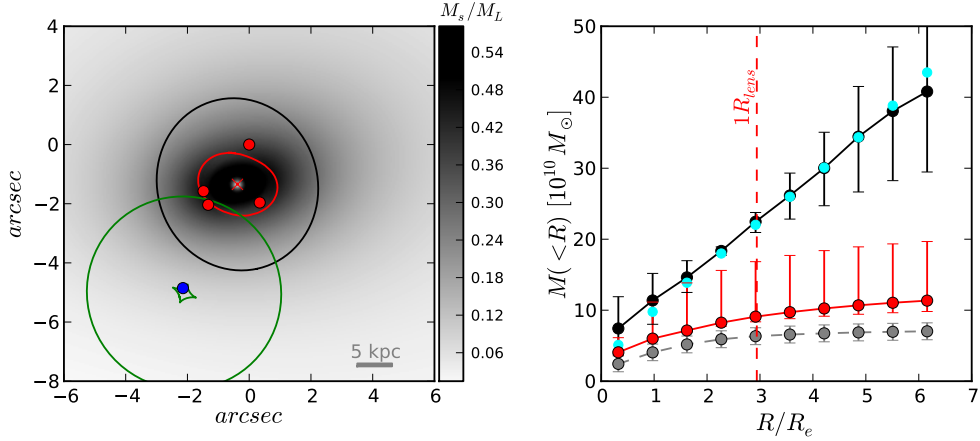}
\caption{Comparison of two different modeling strategies on the lens
  PG1115+080. {\sl Left:} A fit to a three-component lens model
  (stellar, dark matter, and external group) without population
  synthesis.  Red dots mark the image positions and the red curve is
  the model critical curve.  The blue dot is the model source position
  and the green curves show the model caustics.  The grayscale
  indicates stellar-mass fraction, while the black ellipse indicates
  the ellipticity and position angle of the stellar component. The
  semi-major axis of the latter is arbitrary, and set here to $2\rl$.
  {\sl Right:} The red and cyan dots show the stellar and total
  enclosed mass respectively from the model in the left panel.  The
  red error bars enclosing the red dots correspond to a $1\sigma$
  region around the best $\chi^2$.  The gray and black dots with error
  bars are the stellar mass and total enclosed mass respectively from
  the models in \lfsf, which use pixellated lens models and population
  synthesis.\label{fig:lensmodels}}
\end{center}
\end{figure*}

In Section~\ref{sec1} of the present paper we summarize the method of
deriving the dark-matter profile as $\ml-\ms$.  We also consider the
technique of simply fitting a two-component lens model.  The latter
technique, in a test case (see Fig.~\ref{fig:lensmodels}) appears
adequate for estimating $\ml$ but significantly overestimates $\ms$.

In Section~\ref{sec:cm} we fit well-known NFW and Hernquist profiles
to the dark and stellar-mass profiles respectively.
Figure~\ref{fig:paramfit} shows the NFW and Hernquist parameter
estimates and uncertainties for two of the galaxies, while
Figure~\ref{fig:profiles2} shows the dark-matter profiles for the same
two galaxies. The NFW fits automatically provide a virial mass $\mv$
and a concentration $c$, in effect extrapolating the dark-matter
profile out to the virial radius $\rv$.  Figure~\ref{fig:cm} shows
$\mv$ and $c$ for all 18 galaxies.  The trend shown in \cite{B07} is
seen to extend down to virial masses of $10^{12}M_\odot$.  We remark
that the NFW fits shows a characteristic banana-shaped near-degeneracy
between the virial mass $\mv$ and the concentration $c$. These
contribute a spurious inverse correlation between $\mv$ and $c$, but
they are much smaller than the overall trend.

In Section~\ref{sec:am} we use abundance matching
\citep[e.g.,][]{mos10,GUO10} to derive a virial mass $\mvam$ directly
from $\ms$.  The two estimates $\mvam$ and $\mv$ tend to agree in the
majority, but there are cases of strong disagreement.  Interestingly,
the latter are all galaxies in dense environments.  We also consider
the option of constraining the NFW fit such that $\mv=\mvam$.
Figure~\ref{fig:profiles-all} shows how the mass profiles get modified
if this is done, while Figure~\ref{fig:cmam} shows how the $c$-$\mv$
distribution changes.  In the latter case, the scatter increases
considerably.

In Section~\ref{sec:ac} we attempt to reconstruct the initial
concentrations, by fitting the adiabatic-contraction model of
\cite{gnedin04}.  We find that the usual prescriptions for
adiabatic-contraction imply unrealistically low values of $\cinit$,
but weaker adiabatic contractions do fit our results (see
Figure~\ref{fig:ac}). By tweaking the average radius in the
adiabatic-contraction prescription (which can be interpreted as mass
loss during adiabatic contraction) we can obtain agreement with the
data.  The inferred $\cinit$ are shown in Figure~\ref{fig:cmini},
from which it appears that adiabatic contraction can explain part of
the $c$-$\mv$ trend but is unlikely to be the sole origin of it.

\section{Multi-component fitting vs Stellar Population Modelling}
\label{sec1}

The starting point of the present work is the models in \lfsf\ of the
projected stellar and total surface mass density from a sample of
lensing galaxies. We obtained independent maps for the stellar mass
and total mass, using archival data from the CASTLeS
survey\footnote{http://www.cfa.harvard.edu/castles}. The maps of
stellar mass, $\ms$, were derived by fitting stellar population
synthesis models to photometry in two or more bands assuming a 
\citet{chab03} initial mass function (IMF). The total or lens mass,
$\ml$, was mapped by computing pixellated lens models that fitted the
lensed images and (where available) time delays.  Detailed error
estimates were derived in both cases.  The enclosed total mass is well
constrained at projected radii where images are present.  At smaller
and larger radii, $\ml$ becomes progressively more uncertain.  The
outer radius of the mass maps is $2\rl$ where $\rl$ is the radius of
the outermost image.  Since $\rl$ depends on the redshift and details
of the source position, $2\rl$ varies greatly among galaxies --- from
a quarter of the half-light radius ($\re$) to several $\re$.

Of the sample modelled in \lfsf, 18 galaxies are early type systems.
We exclude the Einstein Cross Q2237, which is the bulge of a spiral
galaxy; B1600, which is likely to be a late-type galaxy viewed
edge-on; and B1608, which is an ongoing merger.  For these 18 early
type galaxies, there is no evidence of a significant gaseous
component, and hence we may assume that $\Delta M = \ml-\ms$ is a map
of the dark matter distribution.  The lensing maps tend to have
similar orientation to the stellar mass, and hence the $\Delta M$ maps
are fairly elliptical as well \citep{fe08}.  In \lfsf\ we
obtained enclosed dark matter profiles $\Delta M(<R)$ using a
circularized aperture along the elliptical isophotes, i.e. following
the luminous distribution.

An alternative approach \citep[see e.g.][]{slacs10,trott10}
consists of fitting a parametric lens model
with separate components for stellar and dark matter.  If the stellar
component can be correctly recovered by this method, the analysis
based on stellar population synthesis constrained by multiband
photometry will be dispensable.  The two approaches are contrasted in
Fig.~\ref{fig:lensmodels} in the case of the quad PG1115+080.  On the
one hand we prepared separate models for $\ml$ and $\ms$ --- a
pixellated lens model for $\ml$ and a stellar-population model from
the photometry for $\ms$.  On the other hand, we fitted the lensing
data to a multi-component parametric lens model: a de~Vaucouleurs
profile, plus an NFW halo, together with a singular isothermal sphere,
adding external shear to account for a nearby galaxy group.  We used
the {\tt gravlens} program \citep{gravlens} together with
Markov-chain Monte-Carlo (MCMC) to search for the best fit parameters.  The
effective radius was constrained to lie within the observational
uncertainty $\re=0.85 \pm 0.07$~arcsec \citep{tk02b}.  The positions,
ellipticities and position angle were also allowed to vary.

We see from Fig.~\ref{fig:lensmodels} that the parametric and
pixellated lens models give similar results for the total-mass
profile. The pixellated method provides uncertainty estimates because
it generates an ensemble of models.  It is also computationally
faster. However, the stellar mass is strongly overestimated by the
parametric lens model, compared to the estimates based on population
synthesis.  In other words, the attempt to infer stellar masses from
the lensing data alone fails.  An explanation for this is suggested by
the error bars on the total-mass.  We see that lensing provides a good
estimate of the enclosed mass at radii comparable to the images, but
gets progressively more uncertain as we move inwards or outwards.
Thus, attempting to extract the profile of a sub-component from this
already-uncertain total-mass profile (without adding more data) will
tend to amplify the errors.

Hence, for the rest of this paper, we will use the separate models of
$\ml$ and $\ms$ from \lfsf.

\section{Virial mass and concentration}
\label{sec:cm}

With the maps of $\Delta M=\ml-\ms$ surface mass density in hand, we now proceed to
estimate a virial mass $\mv$ and a concentration $c$ for each of the
lensing galaxies. The method we adopt is to fit the profiles of
$\Delta M(<R)$ to the cumulative projected mass of an NFW profile,
which is given by:
\begin{equation}
  M^{\textrm{\tiny{NFW}}}(<R) = 4 \pi \rho_s r_s^3 \times \mathcal{F}(R,r_s)
\label{eq:nfw}
\end{equation}
where
\begin{equation}
\mathcal{F}(r,r_s) = \ln{\frac{r}{2r_s}} + \left\{ \begin{array}{ll}
\frac{1}{\sqrt{1-(\frac{r}{r_s})^2}} \cosh^{-1}\frac{r_s}{r} & (r < r_s) \\
\stackrel{\phantom{-}}{1} \phantom{\frac{1}{\sqrt{\frac{r}{r_s}^2}}}& (r=r_s)\\
\frac{1}{\sqrt{(\frac{r}{r_s})^2-1}} \cos^{-1}\frac{r_s}{r} & (r > r_s) 
\end{array} \right.
\end{equation}
Here $r_s$ and $\rho_s$ are the scale radius and scale density
parameters on which the NFW profile depends.  For $\Delta M(<R)$ we
assumed an error $\sigma=\sqrt{\delta_{\ml}^2 + \sigma_{\ms}^2}$,
where $\delta_{\ml}$ is half of the $90 \%$ confidence interval given
by the ensemble of lens-mass models, and $\sigma_{\ms}$ is the
standard deviation of stellar mass from population synthesis.  The
best fit values of the parameters can be found in Table \ref{tab2}.

\begin{figure}
\includegraphics[scale=0.63,bb=108 230 523 581]{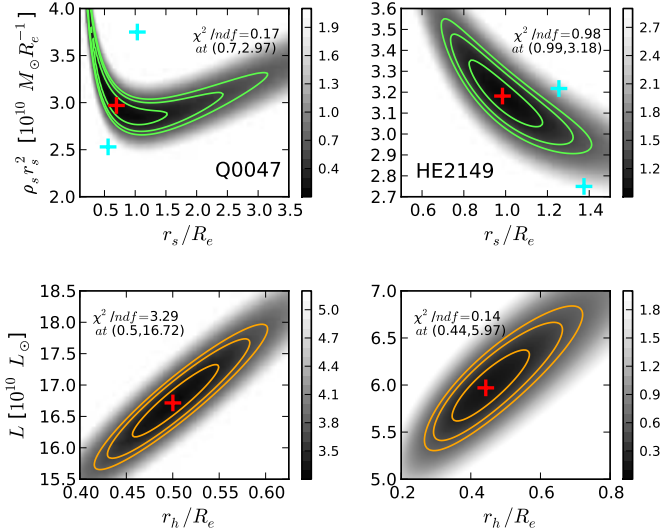}
\caption{Parameter fits to dark matter and stellar ($\Delta M$ and
  $\ms$) profiles for the lens galaxies B0047 and
  HE2149. Top: A $\chi^2$ map of the NFW parameter
  space for $\Delta M$, in grayscale with contours of
  $\Delta\chi^2=1,2,3$.  Red crosses mark the overall best fit.  Cyan
  crosses mark best fits to profiles at the steep and shallow ends of
  the confidence region. Bottom: The same for the Hernquist parameters
  for the luminosity $L$. \label{fig:paramfit}}
\end{figure}

\begin{figure}
\includegraphics[scale=0.60,bb=110 300 523 509]{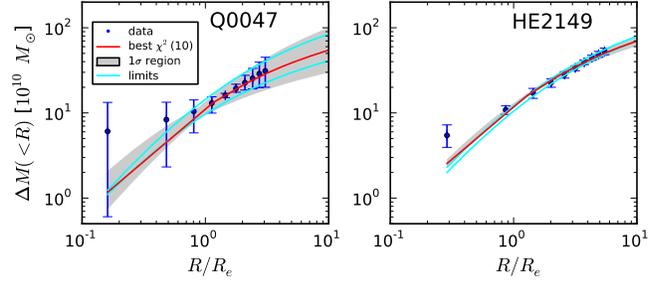}
\caption{Inferred dark-matter profile $\Delta  M$ and NFW fits, for the lens galaxies B0047 and
  HE2149. The red (cyan) line corresponds to the red (cyan) crosses in Figure \ref{fig:paramfit}. \label{fig:profiles2}}
\end{figure}

For two example lenses, Q0047-280 and HE2149-274, we illustrate the
results in more detail in Figs.~\ref{fig:paramfit} and
\ref{fig:profiles2}.  The upper panels of Fig.~\ref{fig:paramfit} show
parameter fits and $\chi^2$ contours. Note that the axes of these two
panels are not simply $r_s$ and $\rho_s$ but rather $r_s/R_e$ and
$\rho_s r_s^2$, in units of $M_\odot \re^{-1}$.  This choice tends to
illustrate the parameters better. To check for the possibility of
multiple local $\chi^2$-minima we generated MCMC chains with $10^5$
steps for each lens.  Such additional minima can be excluded for
physically interesting parameter values. As a further check, we
compute as before the NFW parameters based on a $\chi^2$ search that best fit the steepest and shallowest
profiles $\ml(<R)$ allowed by the \lfsf\ analysis of the lensing data.
These are indicated by cyan crosses in Fig.~\ref{fig:paramfit} and, as
expected, are roughly in the region of the fits to $\Delta M(<R)$.

The lower panels in Fig.~\ref{fig:paramfit} show parameter fits to the
distributions of luminous matter, for the same two example galaxies,
that we will use later, in Section \ref{sec:ac}. We use the well-known profile
of \citet{he90}. The enclosed projected form of the Hernquist (analogous
to Eq.~\ref{eq:nfw} for the NFW) is:
\begin{equation}
L(<R) =
\frac{M}{\Upsilon} \left( \frac{R}{r_h} \right)^2
\frac{\mathcal{X}(R,r_h)-1}{1-\frac{R}{r_h}}
\end{equation}
where
\begin{equation}
\mathcal{X}(r,r_h)= \left\{ \begin{array}{ll}
\frac{1}{\sqrt{1-(r/r_h)^2}}~\mathrm{sech}^{-1} (r/r_h) & r\leq r_h \\
\frac{1}{\sqrt{(r/r_h)^2-1}}~\mathrm{sec}^{-1} (r/r_h) & r\geq r_h 
\end{array} \right.
\end{equation}
Again, the parameters and $1\sigma$ errors as well as our values for
$\rv$ are given in Table~\ref{tab2}. Note in the figure we show the
contours with respect to total luminosity, i.e. $L\equiv M/\Upsilon$.

Figure \ref{fig:profiles2} shows profiles of $\Delta M(<R)$, together
with NFW fits and uncertainties.  There is a tendency for the
innermost point (in these two examples as well in other lenses in our
study) to be higher than the fit.  We note that various simulations
\citep{Moore1998,Navarro2004,Diemand2005} indicate a somewhat steeper
slope than the original NFW.  Recently \cite{Cardone2011} advocated a
generalized NFW profile with an additional parameter. Furthermore, the
presence of baryons will tend to steepen the central dark-matter
profile through adiabatic contraction (although we note that feedback
effects, such as baryon ejecta from supernovae-driven winds, or
dynamical interactions with smaller structures could have the opposite
effect, making the inner dark matter profile shallower). We will
address the issue of adiabatic contraction later, in Section
\ref{sec:am}.  For now we assume that a projected NFW depending on
scale radius $r_s$ and the normalization $\rho_s$ sufficiently
describe the data.

>From the NFW parameters $r_s,\rho_s$ and given the redshift of the
halo, the virial mass $\mv$ and concentration $c$ are easily derived.
Consider first the mass enclosed in a sphere (not to be confused with
the cylindrical enclosed mass Eq. \ref{eq:nfw})
\begin{equation}
\msph(<r)= 4 \pi \rho_s r_s^3 \left\{\ln \left(1+\frac{r}{r_s}\right)-\frac{\frac{r}{r_s}}{1+\frac{r}{r_s}}\right\} \label{eq:msp}
\end{equation}
and the mean enclosed density within a given radius
\begin{equation}
 \langle \rho (<r) \rangle =  \frac{\msph(<r)}{\frac{4}{3}\pi r^3}.
\end{equation}
The virial radius is the $r$ at which the mean enclosed density equals
a certain multiple $\Delta_c$ of the critical density, namely:
\begin{equation}
\langle \rho (<\rv) \rangle=\Delta_c \rho_c(z)
\end{equation}
and the mass within the virial radius
\begin{equation}
\mv=\Delta_c\rho_c(z)\times\frac{4}{3}\pi \rv^3 \label{eq:mvir}         
\end{equation}
is the virial mass.
The concentration is defined as
\begin{equation}
c=\frac{\rv}{r_s}.
\label{eq:cvir}
\end{equation}
The value for the overdensity is
\begin{equation}
\Delta_c = 18 \pi^2 + 82x - 39x^2
\end{equation}
where $x=(\Omega_M(1+z)^3/E(z)^2)-1$ and $E(z)^2 = \Omega_M(1 + z)^3 +
\Omega_\Lambda$ \citep{bn98}.
This value of $\Delta_c$ gives the exact virial radius for a top-hat perturbation
that has just virialized \citep[see e.g.][]{pb80}. The galaxies in our sample would have
virialized well before the observed epoch. Hence, if the observed redshift is
used to derive an $\rv$, the value is unlikely to have the dynamical
interpretation of a virial radius. Nevertheless, since such a definition of
$\rv$ is commonly adopted \citep[e.g.][]{bn98,B07} we adopt it in the present work.

\begin{figure}
\begin{center}\includegraphics[scale=0.64,bb=108 255.5 523 543.5]{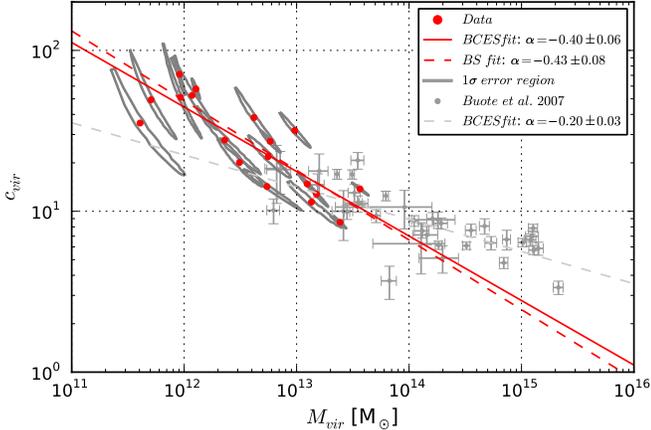}\end{center} 
\caption{Concentration versus virial mass. The red dots represent data from this study. The grey circles shows data from \citet{B07}. The grey contours show the $1\sigma$ error region from Fig.~\ref{fig:lensmodels}. The red dashed line shows a bootstrapping fit to our data. The solid red line shows the result of a bivariate fitting method for correlated errors and instrinsic scatter (BCES) by \citet{ak96} applied to the data. The same method was used by \citet{B07} to obtain the dashed grey line.\label{fig:cm}}
\end{figure}

The values of $c$ and $\mv$, along with errors calculated according to
the projected 1$\sigma$ regions of Figure \ref{fig:paramfit}, are
quantities are listed in Table.~\ref{tab2}.  Figure~\ref{fig:cm} plots
the values --- there have been no previous data showing $\mv$ vs $c$
to such small virial masses.  For comparison, Figure~\ref{fig:cm} also
shows the results from \cite{B07} of the X-ray $c\,$-$\mv$ relation
for 39 systems. These range in $\mv$ from $6\times 10^{12}$ to
$2\times 10^{15}$ $M_\odot$. \cite{B07} fit a power-law
\begin{equation}
c = \frac{c_{14}}{1+z} \left( \frac{\mv}{M_{14}} \right)^\alpha, \label{eq:cm}
\end{equation}
where $M_{14}=10^{14} h^{-1} M_\odot$ is a reference mass and $c_{14}$ and $\alpha$ are constants independent of $M$,
obtaining $\alpha=-0.17\pm 0.03$. Leaving out the $(1+z)$-term gives
$\alpha=-0.20\pm 0.03$.  As in their analysis we use a bivariate
fitting method for correlated errors and intrinsic scatter (BCES) due
to \cite{ak96}, which gives $\alpha=-0.40\pm0.06$.

Note how in Figure~\ref{fig:cm} the projected $1\sigma$ contours cover
a smaller region than simple vertical and horizontal error bars would
imply.  This suggests that more information might be available with
the appropriate tools. Hence, we tried alternative fitting
approaches. 
In a parametric boostrap we randomly resample over arbitrary
points within the $1\sigma$ contours, so that one point per lens is
used for an ordinary least square fit. This is done for $10^4$
realizations. In a later run we ease the requirement of one point per
lens and pick instead a number $n<18$ out of the total number of
lenses to perform the resampling. In all cases the mean value of the
slope $\alpha$ stays the same, but its distribution gets broader for
smaller values of $n$. For $n=18$ the bootstrapping analysis yields
$\alpha=-0.42\pm0.08$. We also employed a piecewise analysis to check how the slope $\alpha$
of the relation evolves and to see whether fits in common mass range
yield similar results. Furthermore we fit a combined sample of 57
objects. The results are shown in Table~\ref{tab1}.

Going from high to low $\mv$ the slope increases from $-0.10\pm0.05$
for $\mv>10^{14} M_\odot$ to $-0.20\pm0.13$ for $\mv<10^{14} M_\odot$
\citep{B07} and finally $-0.40\pm0.06$ for $\mv<4\times 10^{13}$. For
the mass range between $6\times 10^{12}$ $M_\odot$ and $1\times
10^{14}$ $M_\odot$, where the two samples overlap, we do not find
significant differences --- a two-dimensional Kolmogorov-Smirnov test
for the overlapping region gives a $p$-value of $\sim 0.5$ under the
null hypothesis that both samples are drawn from the same population.
However, it should be noted that the reduced sample size and
considerable scatter leads to large errors for both samples.  A trend
of $\alpha$ with virial mass was first suggested by \cite{nfw96} and
confirmed by \cite{BKS01} and \cite{ENS01} for simulations.  Higher
normalization factors compared to simulations are also known from a
lensing study by \cite{CN07}.

\begin{center}
\begin{table*}
\begin{center}
\begin{tabular}{lccccc}
\hline
Sample&Size&Method&$\mv$-range  &  \hspace{0.4cm}$\alpha$ &  $c_{14}$  \\
 & & & [$10^{14} M_\odot$] & &  \\
\hline \hline\vspace{-0.2cm}\\
B07 & 39 & BCES  & $0.06-20$ & $-0.199 \pm 0.026$ & $9.12 \pm 0.43$ \\
B07 & 22 & BCES & $> 1$ & $-0.103 \pm 0.055$ & $7.71 \pm 0.58$\\
B07 & 17 & BCES  & $< 1$ & $-0.201 \pm 0.129$ & $9.46 \pm 2.11$ \\
\hline 
$c$ &18 & BCES & $0.004-0.4$ & $-0.401 \pm 0.064$ & $7.03 \pm 1.49$ \\  
$c$ &18 & BS   & $0.004-0.4$ & $-0.433 \pm 0.078$ & $6.60^{+33.1}_{-6.2}$ \\
$c$ &9  & BCES & $>0.06$ & $-0.203 \pm 0.172$ & $16.98 \pm 12.87$\\ 
\hline 
comb & 57 & BCES  & $0.004-20$ & $-0.278 \pm  0.021$ & $9.62 \pm 0.41$\\
B07$_{0}$ & 39 & BCES  & $0.06-20$ & $-0.172 \pm 0.026$ & $9.0 \pm 0.4$\\ 
$c_{vir,0}$ &18 &  BCES  & $0.004-0.4$ & $-0.381 \pm 0.062$ & $12.02 \pm 2.57$  \\  
CN7$_{0}$ &  62 & N/K & $0.4-100$ & $-0.15 \pm 0.13$ & $10.68 \pm 5.50$\\
\hline
\end{tabular}
\caption{Slope $\alpha$ of the $c-\mv$ relation with uncertainty for
  different samples, sample sizes, fitting methods and mass ranges
  (all errors $1\sigma$, $c_{14}$ errors for BS-method are $68\%$ conf. interterval
  around median). B07 denotes the sample of massive early types in
  \citet{B07}, $c$ stands for the relation as in Fig.~\ref{fig:cm},
  $comb$ gives the combined sample fit consisting of 39 objects of B07
  and 18 lenses of the above $cvir$ sample, CN7 are results from
  \citet{CN07}. The index $0$ to a sample name denotes concentrations
  normalized to $z=0$.\label{tab1}}
\end{center}
\end{table*}
\end{center}

\section{Comparison with abundance matching}
\label{sec:am}
Thus far we have computed $c$ and $\mv$ under the assumptions that
(a)~an NFW profile is a good representation of the dark matter profile
beyond the radial range probed for lens galaxies in \lfsf, and (b)~the
dark matter profile is well constrained by pixellated studies of
stellar and total mass, meaning also that the probed radial range is
sensitive to the scale radius of the NFW profile and that the 
uncertainties give a robust estimate of the suitable mass
distributions.  We now compare the quantities extrapolated to the
virial radius with predictions using both simulations and SDSS
observations.

\begin{figure*}
\includegraphics[scale=0.75]{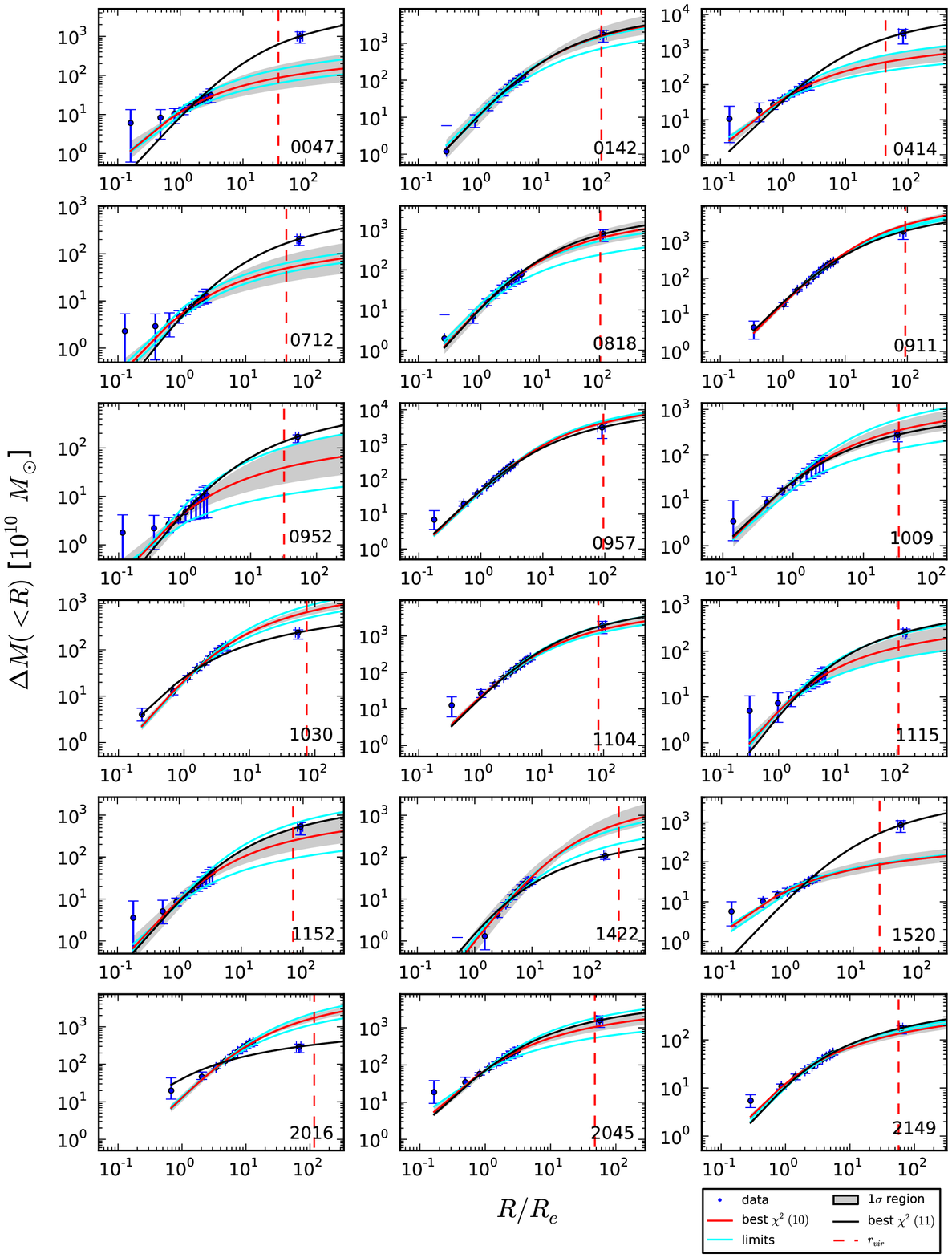}
\caption{Extension of Fig.~\ref{fig:profiles2} to the virial radius
  and for the whole lens sample.  In each panel there are three
  additional elements not present in Fig.~\ref{fig:profiles2}: the
  vertical dashed line marks the the $\rv$ inferred from the NFW fit;
  the outer isolated point shows the virial radius and virial mass
  inferred from abundance matching; the black curve is an NFW fit
  constrained to go through the abundance-matching
  point.\label{fig:profiles-all}}
\end{figure*}

\begin{figure}
\includegraphics[scale=0.64,bb=108 246 523 545]{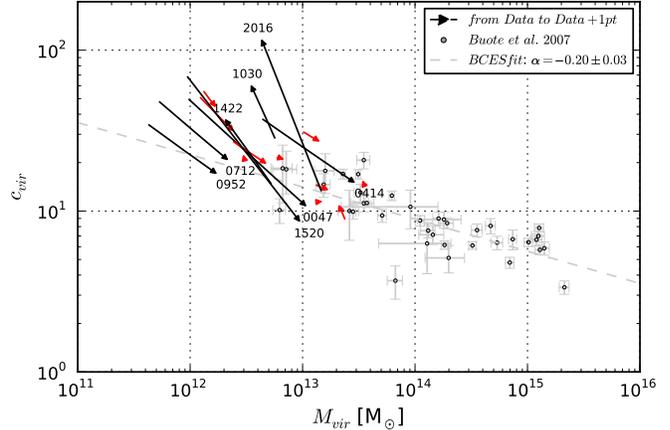} 
\caption{Similar to Fig.~\ref{fig:cm}, but showing how the fits change
  when abundance matching is included. The arrows go from the original
  position (red dots in Fig.~\ref{fig:cm}) to the position for which
  the parameters include information on virial mass from
  \citet{mos10}. The red arrows mark the 10 lenses for which the
  extrapolated NFW analysis and abundance matching give consistent
  answers.\label{fig:cmam}}
\end{figure}

Abundance matching studies like \citet{mos10} and \citet{GUO10} make
use of cosmological simulations and galaxy surveys to determine the
mass dependence of galaxies and their preferred host haloes.  The
stellar mass enclosed within a $2\rl$ aperture is known from our
population synthesis modelling, as shown in \lfsf. The stellar mass
profiles do not change significantly beyond $2\rl$. Thus we use the
$\mh$-to-$\ms$ relation from \cite{mos10} to infer a virial mass and
the corresponding scatter (taken at the $1\sigma$
level). Note that the above abundance matching relations are based on Kroupa/Chabrier IMFs and thus consistent with the stellar masses used here.
Fig.~\ref{fig:profiles-all} shows the extrapolation of the NFW
fits out to the virial radius.  The figure also adds an extra point
$(\rvam,\mvam)$ marking the virial radius and mass from abundance
matching.  For 10 out of the 18 lenses, $\mvam$ turns out to lie within
the $1\sigma$ confidence region (grey shaded) around the original fit
for $\Delta M(R)$.  Accordingly, we show a further NFW fit (black
curve) that is constrained to pass through $(\rvam,\mvam)$.

In Fig.~\ref{fig:cmam} we show how the $c$-$\mv$ scatter plot changes
when we impose abundance matching.  The arrows point from values
without the abundance-matching information to values including it, and
the longer arrows are labelled.  Going from Fig.~\ref{fig:cm} to
Fig.~\ref{fig:cmam} leads mostly to shifts along the direction of the
relation, but the rms scatter with respect to the simple power-law fit
almost doubles, from 0.145 to 0.258.  In comparison, the \citet{B07}
sample has a rms scatter of 0.180. A mildly increased scatter can be
found in simulations by \cite{shaw06} for virial masses between $\sim
3\times 10^{13}$ and $\sim 10^{15} M_\odot$, which is most likely due
to the indistinguishability between substructure and main
haloes. However, this cannot explain the increased scatter we find.
We can conclude that the extrapolation to $\rv$ inferred in Section
\ref{sec:cm} gives a reasonable extension to the $c$-$\mv$ relation.
Abundance matching, on the other hand, appears to introduce a large
discrepancy in some cases. Three of the galaxies (MG2016, B1422 and
B1030) show a shift to a much higher concentration when abundance
matching is imposed.  These are lenses for which $\mvam$ lies
significantly below the extrapolation $\mv$.  For MG2016, $\mvam$ is
even smaller than $\Delta M(<R)$ at the outermost radius of the lens
model. Further three galaxies (Q0047, MG0414, SBS1520) show large
shifts towards lower concentrations. They belong to the highest
redshift galaxies in our sample and are probed in an exceptionally
large radial range, up to $10\%$ of the virial radius (see column
$2\rl/\rv$ in Tab.~\ref{tab2}). Moreover, MG2016 and SBS1520, which
exhibit strongly discrepant $\mvam$, have reconstructed mass profiles
with comparatively small uncertainties.  

So what is the reason for this discrepancy?  A possible explanation is
suggested by a visible correlation between the length of the arrows
and the environment of the lenses.
For extrapolated virial masses much lower than $\mvam$ one may
argue that lens profiles are shallower in group or cluster
environments than in more isolated locations. This reflects the
inverse proportionality of concentration and enclosed mass and is a
consequence of hierarchical structure formation. Extrapolating mass
profiles from shallower profiles leads necessarily to lower masses at
$\rv$. In other words, if the $\mh(\ms)$ obtained from abundance matching is
employed to determine $\mvam$, we implicitly assume an isolated galaxy
located within a "typical" halo with respect to its stellar content and the
halo definition used in the abundance matching procedure. For lenses
with extrapolated virial masses much larger than $\mvam$ the mere
effect of the projected cluster environment might become more
important, that is, although being relatively shallow, the projected
total mass profile is strongly influenced by dark matter in the
cluster acting as an additional convergence. This again causes the
extrapolation to be significantly different from $\mvam$. Examples for
the latter case are MG2016 and B1422, which are located in the densest
environments among our lenses with large groups or clusters showing
many nearby galaxies (cf.~Table~1 in \lfsf).

All 8 lenses for which $\mvam$ is strongly discrepant with $\mv$ are
in dense environments, whereas for 6 out of 10 remaining galaxies,
there are no nearby objects reported so far.  Current
abundance-matching prescriptions do not consider environmental
effects. Our results suggest that environment may significantly
influence the $\mh(\ms)$ function.

\section{Adiabatic contraction}
\label{sec:ac}
In the following section we will assess the extent to which the
$c\,$-$\mv$ relation could be caused by adiabatic contraction of the
halo. \citet{BFFP86} proposed that during the formation of galaxy-sized
structures, the collapse of the dissipative baryons towards the centre
of the forming halo would extert a reaction on the dark matter density
profile, making it steeper. This effect would mean that simple N-body
simulations, such as those that led to the proposal of the NFW profile
\citep{nfw96}, would underestimate the inner slope of the halo.

The concentrations derived in Section~\ref{sec:cm} would therefore
represent the state of the halo after adiabatic contraction (AC).  In
this Section, we will refer to these concentrations as $\cfinal$.
Before AC, the concentrations are thought to have a lower value
$\cinit$.  At present it is not clear whether or not $\cinit$ differed
from $\cfinal$ \citep[e.g.,][]{abadi10} and it is conceivable that the
impact of AC on dark matter profiles might be overestimated by
commonly used recipes for baryonic cooling. 
Additional mechanisms such as dynamical
encounters with smaller structures \citep[see e.g.][]{elz04,cdw11}, or
the ejecta of baryons triggered by supernovae-driven winds
\citep{LAR74,DS86,BGB07} could lead to the opposite effect, making the
inner slope of the dark matter profile less cuspy.
In this paper we only consider the effect from the more fundamental
process of contraction during the formation of the halo.

\begin{figure*}
\begin{center}
\includegraphics[scale=0.75]{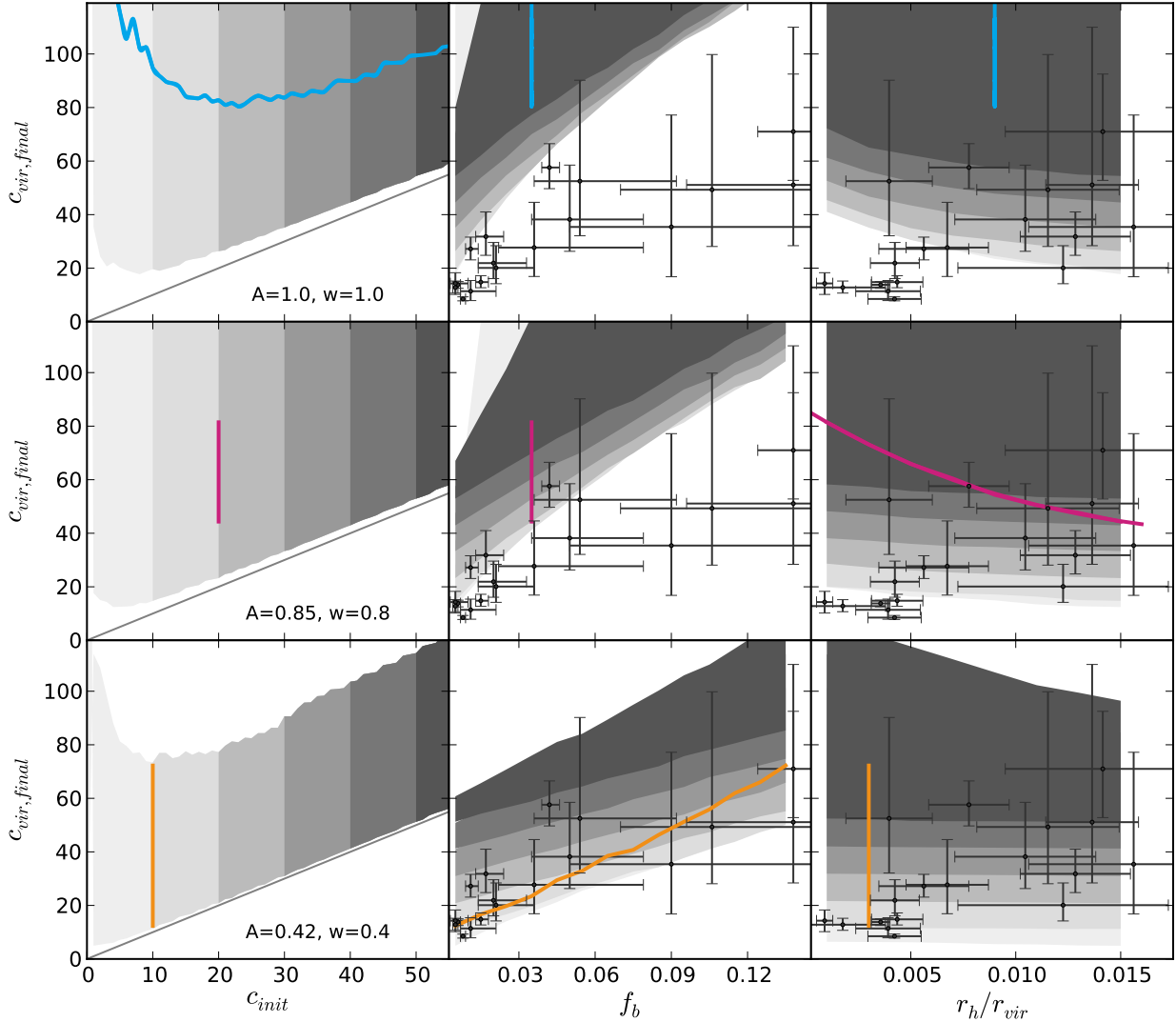}
\caption{Adiabatic contraction following \citet{gnedin04}.  Each row
  corresponds to a different set of values for the parameters $A$ and
  $w$ in Equation~(\ref{eq:fudge}).  Each row shows the final
  concentration against: (left)~the initial concentration,
  (middle)~the baryon fraction, and (right)~the scale radius of the
  baryons.  The initial concentration can be read off the middle and
  right panels by comparing the gray bands with the left panel of the
  same row.  The colored lines illustrate the effect of holding fixed
  two out of (i)~initial concentration, (ii)~baryon faction and
  (iii)~baryon scale radius.\label{fig:ac}}
\end{center}
\end{figure*}

\subsection{Comparing Adiabatic Contraction Models}

To analyze this issue, we make use of the halo contraction program
of \citet{gnedin04}, which computes the change in the dark-matter
density profile under AC, keeping $rM(<r)$ conserved. To take account
of a wide range of orbit eccentricities the code invokes the power-law
\begin{equation}
\bar{r}/\rv = A(r/\rv)^w \label{eq:fudge}
\end{equation}
to describe the mean relation between orbit-averaged and current
radius, and modifies the adiabatic invariant to $rM(<\bar{r})$.
Eq. \ref{eq:fudge} changes the eccentricity distribution of
the mass profile, which is thus distorted by the usage of a mean
radius in the invariant. Parameter $A$ defines the maximum
eccentricity and causes $rM(<\bar{r})$ to be larger than $rM(<r)$ for
$r/\rv<0.44$ and smaller for $r/\rv>0.44$. A larger invariant means
more mass in the center at the expense of the outer parts of the
halo. The parameter $w$ defines how strong the shift is.  The smaller
$w$ the fewer mass at the center.

The case $A=w=1$ therefore corresponds to the original prescription of
\cite{BFFP86}, where the orbits are assumed to be completely circular.
This case can be understood as an upper limit to AC.  The program
provides the necessary resolution for comparison with our data, i.e.,
down to $10^{-3}\rv$. The input parameters are $f_b$, the baryon
fraction enclosed within $\rv$, the baryon scale length and
the initial concentration of the dark matter halo, $\cinit$.  We take
the baryon fraction as $\ms(<2\rl)/\mv$, where $\ms(<2\rl)$ denotes
the stellar mass enclosed in the total reconstructed radial range.
For the baryon scale length we use the fitted Hernquist scale
radius $r_h$ derived in Section~\ref{sec:cm}.  This is preferred to
making use of $\re \approx r_h \times 1.8$ \citep{he90}, because our
measured $\re$ -- derived from the Petrosian radius -- do not agree
precisely with the half-light radius of Hernquist profiles, which is a
consequence of projected radii and circularized mass
profiles. Furthermore, the Hernquist profile is originally used for
the surface brightness distribution, whereas we fit in this case
surface mass profiles. The $r_h/\re$ best-fit values turn out to be
mostly lower but close to $1/1.8$.


We run the contraction routine for a grid of parameters
$(\cinit,f_b,r_h/\rv)$ ranging from $(5,0.005,0.001)$ to
$(60,0.135,0.015)$ in steps of $(1,0.01,0.002)$. We then fitted the
contracted profiles, via Eq.~\ref{eq:msp} to the data $\Delta M(<R)$
for $R/\rv$ ranging from $\sim0.006$ to $0.12$.

\begin{figure}
\includegraphics{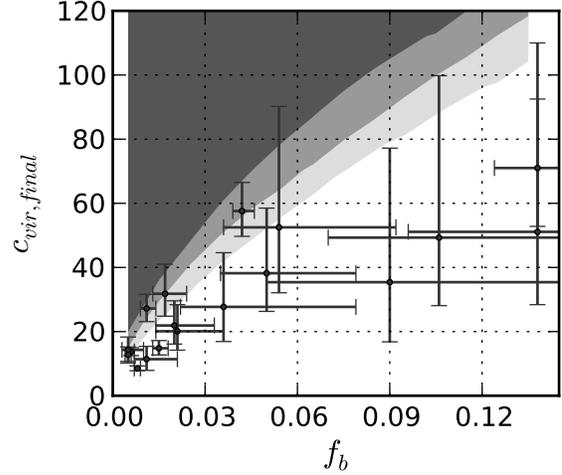}
\caption{Final concentration versus baryon fraction depending on size
  of radial window for the AC prescription of
    \citet{gnedin04}. Bright to dark grey corresponds to aperture
  sizes $(\sim0.003-0.06)\rv$ ,$(\sim0.005-0.09)\rv$ and
  $(\sim0.006-0.12)\rv$. \label{fig:acw}}
\end{figure}

There are a number of uncertainties entering the analysis:
\begin{enumerate}
\item since the radial extent of a reconstructed profile is limited to
  2 times the angular $\rl$ and a finite resolution, the aperture size
  changes from lens to lens,
\item in order to mimic the limited probed range (henceforth
  called aperture) by an equivalent range in the contracted profile,
  $\rl$ must be expressed in units of $\rv$,
\item baryon fraction as well as baryonic scale length depend on $\mv$
  and $\rv$ which are extrapolated quantities with their respective
  uncertainties.
\end{enumerate}

Figure \ref{fig:ac} shows the results for three scenarios of adiabatic
contraction. The leftmost panels show initial versus final
(i.e. contracted) halo concentration.  The top row corresponds to the
original proposal of \citet{BFFP86} ($A=1.0$, $w=1.0$, i.e., no correction
for anisotropic orbits). In this case, we illustrate the increase in
concentration as a blue line for fixed values of $f_b$ and
$r_{h}/\rv$. The growing concentration towards low $\cinit$ is a
consequence of the interplay between radial aperture -- i.e. the
extent of the extrapolation -- and the region affected by AC.  The
smaller the initial concentration the larger $\cfinal/\cinit$ towards
small radii for the same $f_b$ (we refer to this as the
\emph{AC-sensitive case}.).  As $\cinit$ increases the difference
between final and initial profile subsides. Note that in our analysis,
the further out we can probe the halo, the less affected is the fit
and the extrapolation. However, for different combinations of $f_b$
and $r_{h}/\rv$, similar curves fill the grey-shaded region.  To
enable proper differentiation with respect to initial concentration,
we choose different shades of gray. The middle (rightmost) panels show
the final concentration versus baryon fraction (baryon-to-virial
radius fraction). The gray shaded regions map the same areas as those
in the leftmost panels. The black dots with error bars represent our
data. For the \citet{BFFP86} case (top), there is clearly a
disagreement between observationally inferred and contracted
profiles. Especially the low-$f_b$ and low-$r_{h}/\rv$ regions show
significant departure from even the lowest final concentrations of the
generic haloes.  The middle row of Figure \ref{fig:ac} shows the AC
prescription of \citet{gnedin04}, that implements eq. \ref{eq:fudge}
with fiducial values $A=0.85$ and $w=0.8$ to take into account
eccentric orbits. This phenomenologically motivated ansatz leads to
smaller concentrations. There is still significant disagreement
between data and simulated contraction. The behavior of
$c_{vir,final}$ as a function of $r_h/\rv$ for constant
$c_{vir,final}$ and $f_b$ is indicated by the solid magenta line. For
the panels in the bottom row of Fig.~\ref{fig:ac}, we changed the
pre-defined values of $A$ and $w$ to 0.42 and 0.4, respectively. The
orange line shows for fixed $\cinit$ and $r_{h}/\rv$ the final
concentration as a function of $f_b$. These values for $A$ and $w$
give good agreement with the lensing data even for low $\cinit$,
between 1 and 10. Compared to the AC prescriptions shown in the top and
middle rows, the range of final concentrations is narrower,
corresponding to a shallower $\cfinal$-$f_b$ relation (middle
panels). 
The latter can be equivalently expressed in terms of mass not
drawn into the central region $<0.1\rv$.  Comparing mass profiles
contracted according to \cite{gnedin04} with the $(A=0.42,w=0.4)$-case
shows that the latter transports less mass ($\sim 0.4\%$ of the virial
mass) into the central halo region, $<0.1\rv$.\\

One of the intriguing results of this study is that even with a simple
assumption of a common, mass-independent initial concentration, most
of the final concentrations can be explained if ($A$,$w$) are
conveniently adjusted and $r_{h}$ is allowed to vary within
uncertainties. There are a variety of results we summarize in the
following.

\begin{itemize}
\item There is slight evidence for lenses with lower baryon fraction
  to require higher initial concentrations.
\item Both smaller $\cinit$ and smaller $r_h/\rv$ produce steeper
  $\cfinal(f_b)$ curves. This effect is independent of the
  AC-sensitive case at very low $\cinit$ (explained above).
\item When $A$ and $w$ are reduced, $\cfinal(f_b)$ and $\cfinal(r_h)$
  become flatter, i.e. the differently shaded $\cinit$ regions of the
  leftmost panels are mapped to narrower regions in the middle and right
  panels. Moreover their overlap is reduced.
\end{itemize}

To study how sensitive our results are to the radial range, we
additionally provide in Fig.~\ref{fig:acw} the results for reduced
aperture sizes, i.e. $(\sim0.005-0.09)\rv$ and
$(\sim0.003-0.06)\rv$ using the parameters $A=0.85$ and $w=0.8$. 
We find that even for the smallest radial range fits do not yield
agreement with our lens data. The case that we underestimate the
radial extend of our lenses by a factor of two is in light of the
relatively small uncertainties of the mass profiles and the errors
attached to $r_s$ unlikely. Larger radial apertures yield even more
disfavored final concentrations compared to our lens data.

\subsection{Initial Concentration from Weak Adiabatic Contraction}

\begin{figure}
\includegraphics[scale=0.64,bb= 110 246 523 545]{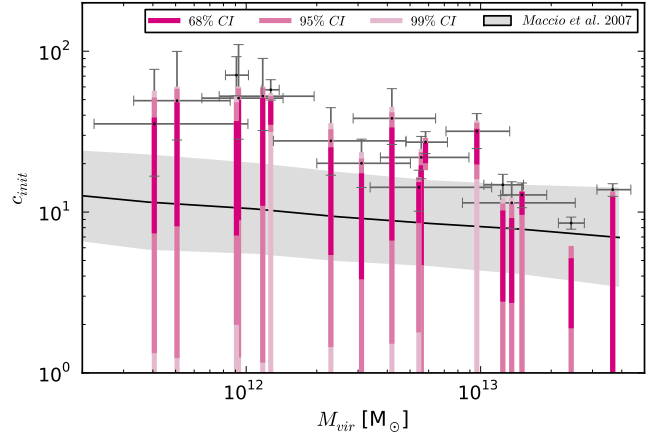}
\caption{As in Fig.~\ref{fig:cm}, but with initial concentrations.
  The colors of the columns from dark to bright correspond to the
  $68\%$, $95\%$ and $99\%$ confidence intervals of a range of
  ($c$,$f_b$,$r_{h}/\rv$) values which produce a $\cfinal$ in
  agreement with our data. For comparison we include
  $\cfinal(\mv)$ and results from simulations by \citet{macc07}.
  The solid line indicates their mean concentration together with
  a $2\sigma$ band (grey region).
  \label{fig:cmini}}
\end{figure}

For the weak AC case $(A=0.42,w=0.4)$ we compare the number of
different ($\cinit$,$r_h/\rv$) combinations producing final
concentrations in agreement with our lens data and infer a
$\cinit$-$\mv$ plot as before, enriched by the information of the
frequency distribution of initial concentrations (see
Fig.~\ref{fig:cmini}).
Although most of the data can be reproduced even by few initial
concentrations of $\sim 1$, most of the ($\cinit$,$r_{h}/\rv$)
combinations with high $\cinit$ produce final concentration in
agreement with $\cfinal$ and $f_b$. The hue of the magenta column
indicates the frequency distribution of $\cinit$ values whereas the
$68\%$ ($99\%$) confidence interval is highlighted by strongest
(faintest) color.

Certainly, no strong quantitative conclusions can yet be drawn from
these results, but judging by the confidence regions a strongly
flattened $\cinit(\mv)$ relation seems likely. These results are in
agreement with the inital halo concentrations estimated by
\citet{lintt06}, who used a simple model of spherical collapse. In
that model, massive galaxies from density fluctuations between 2 and
3$\sigma$ -- roughly mapping the same mass range as our lensing
galaxies -- were found with initial concentrations between 3 and 10,
with the most massive ones having lower concentrations. A flattened
low-mass end of the $c-\mv$ relation is expected by simulations. To
show this we include results from \citet{macc07} (solid line and grey
$2\sigma$ band in Fig.~\ref{fig:cmini}). $c(M,z)$ curves based on a
toy model by \citet{BKS01} for redshifts in a range from 0 to 1.4 are
in good agreement with results from N-body simulations. The toy model
includes the free parameter $K$ which takes into account the
contraction of the inner halo beyond that required by the top-hat
formation scenario. This contraction parameter is fixed for all haloes
in their simulation. The difference between the simulated and observed
$c-\mv$ relation is a well-known issue and matter of ongoing
studies. It is however worthwhile to mention that this discrepancy is
even stronger for low virial mass.  From the comparison between
$\cinit$-$\mv$ found in this study and simulations that investigate
the redshift-dependence, we can conclude that Adiabatic Contraction
alone is not enough to explain the slope of the relation.

\section{Conclusions}
\label{sec:conc}
Strong gravitational lensing on galaxy scales constitutes a powerful tool to
characterize dark matter haloes. In addition, combining photometric
studies with stellar population synthesis allows us to assess the
interplay between baryons and dark matter in the central regions of
galaxies.  This paper extends the work of \lfsf\ by exploring
in detail the concentration of the dark matter haloes of 18 massive
early-type lensing galaxies. On a concentration-virial mass diagram
(Fig.~\ref{fig:cm}) we find these haloes to confirm and extend towards lower
masses the relationship observed in X-rays by \citet{B07}.

Our sample includes information about the baryon fraction, enabling us
to explore the validity of adiabatic contraction prescriptions, such
as the one of \citet{BFFP86} or \citet{gnedin04}. We find that the
standard modelling gives rather high final concentrations compared to
our observations (Fig.~\ref{fig:ac}). 
A tweak of the parameters in the AC prescription of
\citet{gnedin04} cause the gain in mass of the central region
($<0.1\rv$) to be $\sim 4\times 10^{-3} \mv$ lower than in the
previous case, which helps to solve the discrepancy. Furthermore, this
results in a rather flat relationship between initial concentration
(i.e. pre-AC) and halo.

We emphasize that this paper focusses on the fundamental aspect of adiabatic
contraction caused by the collapse of the baryons during the formation
of the haloes. 
Additional mechanisms acting later, arising from dynamical
interactions \citep{elz04,cdw11} or stellar feedback resulting in the
expulsion of baryons \citep[see e.g.][]{rg05,BGB07} may alter the
inner slope of the dark matter halo, although these mechanisms are
expected to be more important in lower mass galaxies.
Nevertheless, the tweak in the AC prescription of \citet{gnedin04}
could be interpreted as one of these mechanisms playing a role in the
evolution of the structure of the haloes. In any case, our analysis
suggests that adiabatic contraction can explain only part of the
$c$-$\mv$ trend but is unlikely to be the sole origin of it.

\section*{Acknowledgements}
We would like to thank the anonymous referee for comments that helped to improve this paper.

\begin{center}
\begin{table*}
\tabcolsep1.2mm
\begin{tabular}{rlccccccccc}
\hline
\textbf{Lens} & $\mathbf{z_L}$ & $\mathbf{r_s}$ & $\mathbf{\rho_s}$ & $\mathbf{\rv}$ &$\mathbf{\frac{2\rl\phantom{\Big{|}}}{\rv}} $& $\mathbf{r_h}$ & $\mathbf{\frac{M}{\Upsilon}}$ & $\mathbf{M_{vir}}$ &  $\mathbf{c_{vir}}$ & $\mathbf{\frac{\ms}{M_{vir}}}$\\
\vspace{0.1cm} &  & $\mathbf{[kpc]}$ & $\mathbf{[10^{8}~\frac{M_\odot}{kpc^3}]}$ & $\mathbf{[kpc]}$ & & $\mathbf{[kpc]}$ & $\mathbf{[10^{10}~L_\odot]}$  &  $\mathbf{[10^{12}~M_{\odot}]}$ &  &\\
\hline \hline\vspace{-0.1cm}\\
\vspace{0.1cm}Q0047&$0.485$&$3.69^{+3.96}_{-2.16}$&$4.22^{+8.67}_{-3.28}$&$188.43^{+29.56}_{-21.33}$&$0.079$&$2.57^{+0.24}_{-0.22}$ & $16.7^{+0.6}_{-0.5}$&$0.93^{+0.51}_{-0.28}$&$51.1^{+58.9}_{-22.7}$ & $0.138^{+0.075}_{-0.042}$\\
\vspace{0.1cm}Q0142&$0.49$&$40.3^{+31.0}_{-14.5}$ &$0.10^{+0.11}_{-0.05}$&$459.55^{+106.30}_{-68.40}$&$0.047$&$1.80^{+0.70}_{-0.54}$& $19.16^{+1.76}_{-1.52}$&$13.57^{+1.18}_{-5.20}$&$11.4^{+4.0}_{-3.5}$&$0.011^{+0.01}_{-0.004}$\\
\vspace{0.1cm}MG0414&$0.96$&$6.45^{+4.30}_{-2.72}$&$4.35^{+8.84}_{-2.63}$&$246.78^{+37.97}_{-29.38}$&$0.085$&$2.58^{+0.87}_{-0.63}$&$ 27.62^{+3.51}_{-2.73}$&$4.18^{+2.24}_{-1.32}$&$38.2^{+20.3}_{-11.9}$&$0.05^{+0.029}_{-0.015}$\\
\vspace{0.1cm}B0712&$0.41$&$3.26^{+3.53}_{-1.86}$ &$3.98^{+13.49}_{-2.64}$&$160.37^{+30.50}_{-21.20}$&$0.054$&$1.85^{+0.52}_{-0.39}$&$ 7.70^{+0.88}_{-0.72} $&$0.51^{+0.35}_{-0.18}$&$49.3^{+50.5}_{-21.2}$&$0.106^{+0.075}_{-0.036}$\\
\vspace{0.1cm}HS0818&$0.39$&$16.5^{+9.4}_{-5.8}$  &$0.44^{+0.51}_{-0.24}$&$360.73^{+60.90}_{-45.20}$&$0.063$&$1.53^{+0.40}_{-0.31} $&$ 13.44^{+0.84}_{-0.76} $&$5.58^{+3.33}_{-1.85}$&$21.9^{+7.7}_{-5.8}$&$0.02^{+0.013}_{-0.006}$\\
\vspace{0.1cm}RXJ0911&$0.769$&$56.8^{+7.9}_{-6.7}$&$0.08^{+0.01}_{-0.01}$&$485.04^{+21.35}_{-19.89}$&$0.067$&$2.05^{+0.70}_{-0.52}$&$ 20.11^{+1.38}_{-1.20} $&$24.34^{+3.36}_{-2.87}$&$8.5^{+0.8}_{-0.7}$&$0.008^{+0.001}_{-0.001}$\\
\vspace{.1cm}BRI0952&$0.632$&$3.75^{+6.92}_{-2.31}$&$2.04^{+2.51}_{-1.73}$&$132.54^{+47.55}_{-23.73}$&$0.065$&$2.07^{+0.40}_{-0.32}$&$6.96^{+0.63}_{-0.51} $&$0.41^{+0.61}_{-0.18}$&$35.4^{+41.8}_{-18.6}$&$0.09^{+0.139}_{-0.04}$\\
\vspace{.1cm}Q0957&$0.356$&$49.9^{+8.36}_{-6.37}$ &$0.13^{+0.04}_{-0.02}$&$687.73^{+42.13}_{-34.61}$&$0.074$&$2.45^{+0.29}_{-0.26}$&$26.74^{+0.92}_{-0.84} $&$36.59^{+7.15}_{-5.25}$&$13.8^{+1.2}_{-1.3}$&$0.006^{+0.001}_{-0.001}$\\
\vspace{0.1cm}LBQS1009&$0.88$&$11.5^{+7.4}_{-4.4}$&$0.75^{+0.97}_{-0.54}$&$231.71^{+40.24}_{-31.54}$&$0.079$&$2.84^{+1.24}_{-0.91}$&$8.73^{+1.26}_{-1.02} $&$3.10^{+1.91}_{-1.10}$&$20.1^{+8.3}_{-5.9}$&$0.021^{+0.015}_{-0.007}$\\
\vspace{0.1cm}B1030&$0.6$&$12.0^{+3.2}_{-2.3}$    &$1.11^{+0.53}_{-0.35}$&$327.36^{+24.32}_{-20.46}$&$0.055$&$1.84^{+0.79}_{-0.59}$&$7.28^{+0.80}_{-0.65} $&$5.82^{+1.40}_{-1.02}$&$27.2^{+4.4}_{-4.1}$&$0.011^{+0.003}_{-0.002}$\\
\vspace{0.1cm}HE1104&$0.73$&$26.7^{+6.5}_{-5.0}$  &$0.28^{+0.12}_{-0.08}$&$395.14^{+27.71}_{-23.76}$&$0.075$&$1.72^{+0.53}_{-0.41} $&$ 19.73^{+1.10}_{-1.00} $&$12.43^{+2.80}_{-2.11}$&$14.8^{+2.4}_{-2.1}$&$0.015^{+0.003}_{-0.002}$\\
\vspace{0.1cm}PG1115&$0.31$&$4.27^{+3.99}_{-2.09}$&$3.76^{+12.04}_{-2.65}$&$224.36^{+41.10}_{-29.80}$&$0.055$&$0.89^{+0.53}_{-0.35}$&$7.78^{+0.99}_{-0.77} $&$1.18^{+0.77}_{-0.41}$&$52.5^{+37.7}_{-20.4}$&$0.054^{+0.038}_{-0.018}$\\
\vspace{0.1cm}B1152&$0.439$&$9.43^{+10.29}_{-4.57}$&$0.87^{+2.11}_{-0.55}$&$261.25^{+72.93}_{-44.80}$&$0.047$&$1.76^{+0.36}_{-0.29}$&$12.09^{+0.81}_{-0.67} $&$2.30^{+2.51}_{-0.99}$&$27.7^{+16.9}_{-10.8}$&$0.036^{+0.043}_{-0.014}$\\
\vspace{0.1cm}B1422 &$0.337$&$25.9^{+20.3}_{-8.5}$&$0.14^{+0.13}_{-0.04}$&$368.43^{+99.18}_{-54.28}$&$0.028$&$0.33^{+0.14}_{-0.11}$&$3.63^{+0.20}_{-0.20} $&$5.45^{+5.69}_{-2.07}$&$14.3^{+4.0}_{-4.1}$&$0.005^{+0.005}_{-0.002}$\\
\vspace{0.1cm}SBS1520&$0.71$&$2.35^{+0.93}_{-0.60}$&$14.57^{+3.22}_{-7.67}$&$166.82^{+6.69}_{-5.89}$&$0.101$&$2.36^{+0.90}_{-0.64}$&$14.75^{+1.80}_{-1.40} $&$0.91^{+0.11}_{-0.09}$&$71.0^{+21.5}_{-18.2}$&$0.138^{+0.017}_{-0.014}$\\
\vspace{0.1cm}MG2016&$1.01$&$28.9^{+8.6}_{-6.1}$  &$0.29^{+0.15}_{-0.10}$&$369.53^{+31.19}_{-24.92}$&$0.105$&$0.65^{+0.94}_{-0.56} $&$ 7.67^{+0.94}_{-0.77} $&$15.01^{+4.13}_{-2.84}$&$12.8^{+2.4}_{-2.1}$&$0.005^{+0.001}_{-0.001}$\\
\vspace{0.1cm}B2045&$0.87$& $10.7^{+4.6}_{-3.1}$  &$2.45^{+2.32}_{-1.10}$&$339.55^{+38.59}_{-32.61}$&$0.062$&$4.36^{+0.85}_{-0.68} $&$21.80^{+1.76}_{-1.44} $&$9.63^{+3.67}_{-2.52}$&$31.8^{+9.2}_{-7.0}$&$0.017^{+0.007}_{-0.004}$\\
\vspace{0.1cm}HE2149&$0.603$&$3.42^{+0.65}_{-0.53}$&$7.86^{+3.72}_{-2.53}$&$196.91^{+5.53}_{-5.54}$&$0.091$&$1.53^{+0.42}_{-0.33} $&$ 5.97^{+0.40}_{-0.34} $&$1.27^{+0.11}_{-0.10}$&$57.6^{+8.9}_{-7.9}$&$0.042^{+0.004}_{-0.003}$\\
\hline
\end{tabular}
\caption{Lens, redshift of the lens $z_L$, NFW scale radius $r_s$, NFW
  scale density $\rho_s$, inferred virial radius $\rv$, outermost
  radius of the mass profiles in terms of virial radius $2\rl/\rv$
  (for the innermost radius multiply by $1/19$), Hernquist scale
  radius $r_h$, Hernquist scale luminosity $\ms/\Upsilon$, the virial
  mass $\mv$ as defined in Equation \ref{eq:mvir}, the concentration
  as defined in Equation \ref{eq:cvir}.\label{tab2}}
\end{table*}
\end{center}

\bibliographystyle{mn2e}

\label{lastpage}

\end{document}